\documentclass[twocolumn,twoside,slac]{revtex4}
\usepackage{graphicx}
\usepackage{fancyhdr}
\begin{document}
 {YITP-SB-05-05}

\title{Newtonian Adiabatics Unified}
\author{Alfred Scharff Goldhaber\footnote{goldhab@insti.physics.sunysb.edu}}
\affiliation{C.N.Yang Institute for Theoretical Physics, State University of New York, Stony Brook, NY 11794-3840 USA}

\begin{abstract}
  Newtonian adiabatics is the consistent truncation  of the adiabatic approximation to  second order in small  velocities.  To be complete it must unify two hitherto  disjoint intellectual streams in the study of adiabatic motion.  The newer stream focuses on Berry's induced vector potential, or geometric magnetism, and Provost and Vall\'ee's induced scalar potential, reflecting geometry in Hilbert space.  The older stream focuses on Inglis' induced inertia, influencing the geometry of adiabatic-parameter space. Starting with the Hamiltonian  of the newer stream, unification is simple:   A naive or primitive inertia, whose inverse appears in two terms of that Hamiltonian, is replaced by the convention-independent sum of primitive and induced inertia tensors.
  

\end{abstract}

\maketitle

\section{Introduction -- Newtonian `toy models'}

The dynamics of electrically charged particles interacting through the electromagnetic field has a natural expansion in the velocities of the particles. An all-order expansion in velocities is at best asymptotic in character, because already at third order in velocity one has radiative processes implying dissipation of the purely particle energies, and hence a non-closed system.   Nevertheless, at second order in velocity there is a consistent truncation of the dynamics involving only the particle degrees of freedom.   The electric interactions among particles are given by static Coulomb potentials, while the magnetic interactions are given by a less-familiar form, the Darwin Lagrangian \cite{darwin}.

 This truncated theory is naturally described as Newtonian electrodynamics, involving as it does kinetic energies quadratic in velocities, and interactions among particles which are instantaneous, so that there is no place for retardation or radiation.\footnote{
In the Darwin Lagrangian, among all terms second order in ratios of velocities to the speed of light c is a contribution fourth order in velocity, coming from the velocity-dependence of the inertial mass of each particle.  As observed by Coleman and Van Vleck  \cite{darwin}, this contribution is essential for a consistent description of the relative motion between a current-loop magnet and an electric charge. Thus we have a demonstration by example that, in cases where there is more than one scale for measuring velocities, a second-order truncation in velocity might not be enough to give a description.  This does not contradict the  main points of the present paper, that accounting for terms up to second order is at least necessary for consistent truncation, and that there is a systematic way to obtain the unique form of this truncation.}    In modern parlance, such a theory might be called a `toy model', because important features of the full dynamics still need to be included.  Nevertheless, literally for centuries  Newtonian theory 
was a cornucopia of powerful developments in physics, and even today is the basis for presentations of mechanics in introductory physics courses.  Thus, this is a toy with great value and useful applications.  It is true that for accelerating  charged particles there always will be some radiation.  
  Even for electrically neutral objects interacting through gravity, there also  is inevitable, if unobservably small, radiation. Still, the Newtonian approximation for gravitational systems has proven an enormously rich framework, eminently justifying its continued use even though we know it  is incomplete (not only omitting radiation but also other relativistic effects such as the Einstein contribution to the precession of the perihelion of Mercury).  

The picture seems  quite similar for the case of the adiabatic approximation, giving an effective action for slow degrees of freedom after `integrating out' fast degrees of freedom.
In the same sense as for electrodynamics, stopping the adiabatic expansion at second order gives a consistent truncation.
Of course, as the terms in the expansion are obtained by perturbation in the velocity, one is entitled to the position that even an internally consistent truncation is logically unjustified, because the perturbation expansion has no finite stopping point.  However, the beauty and simplicity of the truncation are so appealing that its internal consistency seems a more than adequate reason to consider it separately.  Newtonian dynamics, including Newtonian electrodynamics, provides an alluring indicator of the potential value in such an approach.  The main point of the present paper is to provide a complete second-order truncation, because different works in the literature omit one or another part.  Let us begin by enumerating those parts (all having geometric interpretations) which go beyond the original Born-Oppenheimer approximation \cite{BO}.

\section{Geometries of Newtonian adiabatics}
Geometry intertwined with dynamics is a pervasive theme in modern physics:  General relativity identifies gravity with the geometry
of spacetime.  Electrodynamics and other gauge theories, as seen in the
context of quantum mechanics, are related to the differential geometry of
a map between points in spacetime and directions in an abstract space.
In this approach, a vector 
potential is seen as a connection characterizing how the map rotates under
infinitesimal motions in spacetime, and the corresponding field strength
is simply the curvature of that connection.  A second theme receiving
continually increasing recognition is the importance of approximation
schemes based on averaging over fast degrees of freedom to obtain the
dynamics for any remaining slow degrees of freedom.  In the context
of nonrelativistic physics this usually is described as an adiabatic
approximation, while in the context of relativistic quantum field theory
the more common label would be by the result, called an 
effective field theory.  Note that effective field theories generally have actions quadratic in time derivatives of the fields, and thus are examples of the Newtonian truncation in the sense used here, even though of course they are fully relativistic.

A  striking connection between the themes of geometry and
adiabatics is Berry's 
discovery \cite{mb1} that adiabatic variation of parameters in a Hamiltonian 
induces effective vector potentials appearing in the kinetic momenta conjugate to such
parameters or coordinates.  Because the structure of the 
parameter space determines the effective vector potentials and resulting
effective magnetic fields, Berry describes the phenomenon as `geometric
magnetism'.  From the perspective of the previous paragraph, it also would 
be reasonable to use the term  `induced (gauge) geometry', as any gauge
interaction may be interpreted geometrically.  

There is still
another kind of geometry
found by Provost and Vall\' ee
\cite{scalar} shortly before Berry's work:  In addition to the vector potential, there
is a scalar potential, which also expresses a geometric structure,
\begin{equation}
\Phi=\hbar^{2}Q_{ij} g_{ij}/2, \label{PV}
\end{equation}
  where in the Hamiltonian
for the slow variables the kinetic term is $K = P_iQ_{ij}P_j/2$.  The `metric'
$g_{ij}$ measures the infinitesimal distance 
(in Hilbert space) between instantaneous 
fast-variable eigenstates corresponding to an infinitesimal change in the
values of the adiabatic parameters.  As such, $g_{ij}$ of course is intrinsically
positive, as is the inverse inertia factor $Q_{ij}$ which multiplies
it, so that $\Phi$ itself always is positive.  A classical interpretation of this potential was given by Aharonov and Stern \cite{AS} for the case of a particle with spin and magnetic moment passing through a region in which the magnetic field varies slowly in direction, allowing application of the adiabatic approximation.  The scalar potential comes from mean-square oscillation of a component of the spin perpendicular to the magnetic field direction.  The reason that in the original discussion \cite{scalar} this term vanishes with $\hbar$ is that the spin is assumed to be aligned along the magnetic field as well as quantum mechanics can allow, so that the mean-square perpendicular components of the spin are proportional to $\hbar$, and would disappear in the classical limit.  An amusing technical point:  In this example, the second factor $\hbar$ in (\ref{PV}) is compensated by a large magnetic quantum number to give a nonvanishing classical spin.

The existence of these beautiful if exotic 
geometrical structures raises
the question whether adiabaticity generically induces or perhaps modifies
more conventional geometry, namely that of the space of slow
parameters.  This space is analogous to
 the space of possible locations of 
a particle, in which geodesic paths are the trajectories
followed if no explicit forces are acting.  
In other words, the metric is given by the inertia tensor for
the slow parameters.  There are two key aspects of this inertia.  First,
it must be large, so that motion is slow enough to make
the adiabatic approximation accurate, but
not so large that the effects of adiabatically induced forces are negligible. 
Secondly, the large inertia may be primitive, i.e., associated with explicit
degrees of freedom in the full action, or induced, i.e., a consequence of the
velocity-dependent coupling associated with the adiabatic variation of 
parameters.

It will be seen a little later that at least one prominent
case of the latter type has
been known for decades.  Nevertheless, the simplicity, universality,
and especially the geometry associated with induced 
inertia seem yet to be accorded
the wide recognition they deserve.

To compute induced inertia, we need to consider systematically contributions 
to the energy
through second order
 in the velocity of slow coordinates, i.e., beyond what 
is needed for the
scalar potential (zeroth order in velocity)
or the vector potential (first order in velocity, though locally ambiguous
because of gauge freedom).
Let us examine a little more carefully
the orders in small parameters of the
relevant geometric contributions to the Hamiltonian.  Berry \cite{mb2}
considered the limit $|{\bf V}|T$ fixed, $T \rightarrow \infty $,
where $T$ is the time for completion of a cycle in parameter space.
However, one may also take $T$ fixed and finite, so that the
area enclosed by the cyclic orbit becomes small in the limit
of small velocity.
  In that
case, assuming that the fast variables (such as a 
large but slowly precessing spin)
are of macroscopic or classical magnitude, 
it is straightforward to show that the (quantum) scalar potential
contribution to the action
is $\propto\hbar^2 T$, that of the induced Berry 
flux is 
$\propto V^2T^2$, and the quadratic contribution to be discussed below
is  $\propto V^2T$.
Thus for fixed $T$ the inertial term and the Berry term are comparable,
and clearly both should be included in a consistent scheme    Clearly if the
fast variables are quantum in scale then all three terms should be taken
into account.

Second-order terms in velocity
have the same form as conventional kinetic energies, 
so that if the slow variables specify coordinates of a
massive particle there already is such a term present.
If there is no such primitive quadratic term, but one wishes to identify
the slow parameters as collective variables, then it is essential to obtain
from the adiabatic evolution itself precisely such a kinetic term.  Even if
there were a primitive contribution, one should expect it to be supplemented
by an induced contribution.  

Berry \cite{mb2} discussed the 
systematic expansion of the total phase associated with an arbitrarily
slow cyclic motion in powers of the velocity.  He observed that, unlike
the case of ordinary time-independent perturbation theory for a finite
system, the adiabatic expansion is an asymptotic series, rather than
a Taylor series with a finite radius of convergence:
There is an exponentially small probability of non-adiabatic jumps, and
this implies an essential singularity at zero velocity.  Nevertheless,
for sufficiently small velocity the first few terms of the series can give
an accurate description of the evolution. These considerations imply that
for a self-contained dynamics one at least
should go to second order in the expansion,
so as to determine completely the inertia 
tensor of the slow degrees of
freedom:  The inertia is a prerequisite for obtaining observable
consequences from the vector and scalar potentials.  

\section{Induced inertia, the final piece in Newtonian adiabatics}
Consider the general problem specified by a time-dependent Hamiltonian,
\begin{equation}
H(t)|\psi\rangle = i\hbar (d|\psi \rangle \! / \!dt) \ \ .
\end{equation}
  If the rate of change for $H$ is
slow (and its eigenvalues do not change), then
in the vicinity of any time $t_0$ we may write 
\begin{equation}
|\psi(t)\rangle= U(t)|\psi '(t)\rangle,
\end{equation}
 where 
one has by definition $U(t_0)=1$, and to first order in velocity $U^
{\dagger} H U$ is
time-independent.  This gives a familiar time-independent perturbation
theory problem to determine $|\psi '(t_0)\rangle$.  The 
equivalent `perturbed' Hamiltonian
is 
\begin{equation}
H' = H(t_0) + {\bf V \cdot P},
\end{equation}
 where the matrix elements of the 
operators $P_i$ are defined by 
\begin{equation}
\langle m|P_i|n\rangle = -i\hbar \langle m|\partial_{X^i}|n\rangle \ \ ,
\end{equation}
with $n \ne m$, and $V^i = \partial_t {X^i}$, the (slow) velocity
of motion in the space of parameters $X^i$.

To first order in ${\bf V}$, the wave function is given by 
\begin{equation}
|\psi(t_0)\rangle = |\psi '(t_o)\rangle  = |n\rangle + \Sigma \alpha_m |m\rangle \ \ ,
\end{equation}
with again $m \ne n$, and 
\begin{equation}
\alpha_m = {\bf V} \cdot \langle m|{\bf P}|n\rangle 
/(E_n - E_m).
\end{equation}
  This means that the instantaneous eigenfunction to
first order in ${\bf V}$ is {\bf not} simply the eigenfunction
of the instantaneous Hamiltonian. What we want to know is the shift in
energy to second order in ${\bf V}$ implied by this shift in the
wave function.  We have arrived at the crucial juncture in the
calculation.  Although $\psi$ is not an eigenstate of $H,$ it is
$H$ which appears in the Schr\"odinger equation, and therefore 
the desired energy must be computed
from the expectation value $<\psi |H(t_0)|\psi >$:
\begin{equation}
\Delta E_n = \Sigma |\alpha_m|^2
(E_m-E_n),
\end{equation}
 which evidently is positive if $|n \rangle$ is the ground state
with respect to the fast variables.  Using the definition of $\alpha_m$, one
may rewrite the energy shift in a suggestive form:
\begin{equation}
\Delta E_n =
\Sigma_{m} |\langle m|{\bf V \cdot P}|n\rangle |^2/(E_m-E_n).  \label{inertia}
\end{equation}
  This should look very
familiar, as it differs only in sign from the well-known
expression for the second-order energy shift in conventional
time-independent perturbation theory.  Just as the negative sign in
the latter case may be understood as a consequence of level repulsion
by mixing potentials, so the positive sign here makes excellent
physical sense:  If one `wobbles' the slow parameters for a system
in its instantaneous ground state with respect to fast variables, that
wobbling can only raise the energy.  It might be interesting to
study the relationship between the different behaviors
for time-independent perturbation theory and adiabatic perturbation
theory of the shifts in
neighboring energy levels (repulsive or attractive) and the behaviors
of the corresponding series (convergent or divergent).

Let us rewrite the expression one more time, as 
\begin{equation}
\Delta E_n =
{\cal I}_{ij}V^iV^j/2,
\end{equation}
 where this implies 
\begin{equation}
{\cal I}_{ij} =
2 Re \Sigma_{m,m \ne n} 
\langle n|P_i|m\rangle\langle m|P_j|n\rangle /(E_m-E_n). \label{inertia'}
\end{equation}
  The inertia tensor 
 ${\cal I}_{ij}$
plays the role of a metric in the space of coordinates $X^i$, as
the principle of least action implies that in
the absence of explicit forces the motion follows a geodesic
path as determined by ${\cal I}$. Of course, if there were also a primitive
quadratic term in the velocities, then it would be the sum of the 
primitive and the induced contributions to the inertia
which would constitute the metric.  Equation (11) represents the
key result.  It implies, as asserted earlier, that for motion of an
instantaneous ground state the inertia tensor or spatial metric receives
an intrinsically positive contribution.  Near a crossing point of two
instantaneous energy levels, where of course the adiabatic approximation
must fail, the Berry vector potential diverges as the inverse first power
of distance from the crossing, while the scalar potential diverges as the
inverse second power \cite{berlim}.  
Because of the extra energy denominator, the inertia
tensor diverges as the inverse third power (slowing the response to applied forces). 

All these effects combine
to protect the ground state from too close an approach to any such crossing,
giving a self-enforcement of the adiabatic approximation. 
On the other hand, for the higher of two states near a level crossing,
the vector and scalar potentials continue to give positive or
repulsive
$1/r^2$ effects, but 
 the induced contribution to the
inertia now is negative, by itself
generating what with repulsive forces becomes an acceleration towards
the 
level crossing, and therefore a possibility of breakdown rather than
preservation of adiabaticity.

The above discussion is quantum-mechanical, whether the slow
variables
are collective or are those of massive `elementary' particles.  
When adiabatic motion is
associated with classical collective variables, 
for example, 
degrees of freedom characterizing a soliton configuration
of classical fields, then there is a well-known procedure for 
computing the kinetic energy in terms of the classical action for
the fields, and identifying this kinetic energy as a quadratic form in
the time derivatives of the collective coordinates \cite{soliton}.  
This gives a
nice continuity between quantum and classical treatments of such 
phenomena.  In both regimes of course the inertia is intrinsically positive
if the associated structure for zero velocity is stable.

An illustration of the quantum procedure for the case of 
collective coordinates
 is the Inglis cranking model,
introduced to describe the
low-lying rotational bands in deformed nuclei \cite{inglis1}.  
The general formula (11) was evaluated for slow rotation of
the symmetry axis of a
spheroidal harmonic oscillator potential containing a Fermi gas of nucleons, with the result
that the moment of inertia takes its rigid-body value.  Later work
on the collective model of nuclei introduced an attractive pairing
force between nucleons, yielding substantially lower and more 
phenomenologically acceptable
values of this inertia \cite {MV60}.
A
systematic algebraic formulation 
of the Inglis cranking model was described by
Lipkin,de Shalit, and Talmi \cite{LST}, 
who obtained a refinement taking account of 
the
`center-of-mass' correction -- The orientation of the nuclear deformation axis is
redundant with the full set of coordinates of all the individual nucleons.
This of course becomes irrelevant if the
slowly varying coordinate is associated with an elementary particle of
large mass.

A case of the
latter  sort was treated  by Littlejohn and Weigert [LW] \cite{lw}, who pursued further the 
considerations of Aharonov and Stern \cite{AS} on a neutral particle with spin (and parallel magnetic moment)
moving through a region in which a strong magnetic field varies slowly both
in magnitude and direction.    LW found a term in the energy proportional to
the square of the momentum, in addition to the usual kinetic energy of the
massive particle.  Thus the kinetic energy is changed, though
only slightly, from the
case without the variable field.  Let us use (11) to obtain the
LW result. In terms of the particle coordinates, the operator we need is
\begin{equation}
\delta H = {\bf (V \cdot \nabla) \hat B \cdot S },
\end{equation}
 where 
${\bf V}$ 
is
the particle velocity, $\hat B$ is a unit vector in the direction of the
magnetic field ${\bf B}$, and ${\bf S}$ is the particle spin. Substituting
into the formula (\ref{inertia}), for a state labeled by spin projection $m$ onto
the direction of ${\bf B}$, we obtain 

\bigskip

$\ \ \ \ \ \ \ \ \ \ \Delta E(m) = {\bf[ (V \cdot
\nabla)\hat B }]^2 (1/gB) \ \  \times $
\begin{equation}
[ |\langle m-1|S_x|m\rangle |^2 - |\langle m+1|S_x|m\rangle |^2] \ \ ,
\end{equation} 
where the
bracket has the value $\hbar^{2}m/2$, and $-gBm$ is the interaction energy of the
spin with the magnetic field.  This expression is identical to that obtained
by LW, except for the sign.   In their analysis, the sign of the
extra term is negative for positive $m$.  That apparent discrepancy 
has a trivial explanation:  Their expansion uses momentum rather than 
velocity, and because the mass appears in the denominator when kinetic
energy is expressed in terms of momentum, an increase in effective mass 
becomes a negative contribution to the energy expressed in terms of 
momentum.  They, like Berry in his discussion of asymptotic 
expansions in powers of the velocity \cite{mb2}, do not discuss explicitly the
significance of the sign of the quadratic energy term.  Therefore, we
may consider the argument here as explaining in terms of basic 
principles  a sign which
was an issue of no special concern in their work. 

 For Inglis of course
the sign was crucial, as a net negative moment of inertia yields
an instability against increase of angular momentum, and is physically
unacceptable as well as clearly unrelated to experiment.  Thus he
obtained the correct sign because he knew what it should be, and tacitly reversed the sign of  
the standard, negative, stationary-state second-order perturbation
energy.  The arbitrariness was noted and corrected afterwards, in a
manner outlined by Goeppert-Mayer [\cite{inglis2}.\footnote{Actually the result mentioned earlier of a rigid-body value for the inertia only was worked out explicitly in \cite {inglis2}.}  This discussion makes clear that
the `new term' of LW represents an independent discovery of induced inertia,
nearly 40 years after it was introduced by Inglis.  Perhaps because they did not identify this effect as induced inertia, they did not use it also to modify the Provost-Vall\'ee scalar potential, as is advocated in the next section of the present paper.

A simple application of induced inertia comes from the almost
trivial problem of the free motion of a hydrogen atom.  By Galilean
invariance, the kinetic energy is $K=(M+m)V^2/2$, where the two masses
are those of the proton and the electron, respectively.  In the 
adiabatic formulation, the first term is primitive, and the second
must be induced by the motion of the center of the Coulomb potential
influencing the electron.  According to the general formula (\ref{inertia'}), this
gives

\vspace{.05in}

\centerline{$m\delta_{ij} = $}
\vspace{-.1in}

\begin{equation}
2\hbar^{2}Re \Sigma_m
\langle -i\partial_{X^i} (n)
|m\rangle \langle m|\! -  \! i\partial_{X^j}|n\rangle /(E_m-E_n),
\end{equation} 
where
$X_i$ is the proton coordinate.  Because of translation invariance,
we may substitute for the gradient with respect to the proton
coordinate the negative gradient with respect to the electron
coordinate.  Consequently, with a little rearrangement the relation
may be expressed in the form 
\begin{equation}
\delta_{ij} = (2m/\hbar^{2}) \Sigma_m\rangle \langle n|x^i|m\rangle \langle m|x^j|n\rangle (E_m-E_n).
\end{equation}

This is nothing but the well-known Thomas-Reiche-Kuhn energy-weighted
sum
rule for electric dipole transitions, the ancestor of a host of sum
rules
extending all the way to high energy physics in the 
the analysis of phenomena such as deep
inelastic
lepton scattering.  The TRK sum rule is
easily derived by elementary commutation relations
of
the position coordinate operator with the Hamiltonian and the
momentum.  For an atom with $Z$ electrons, the left hand side of the
sum rule would be multiplied by $Z$, so that it counts the number of constituents of
the atom contributing to photo-excitation \cite {PT62}.

The analysis presented above indeed rounds out the
picture of induced geometry associated with adiabatic interactions,
adding to gauge geometry and Hilbert space geometry
the even more venerable geometry of ordinary coordinate space.  
In general all these geometrical effects may appear in
any system.  They all do in the LW case, but
even models of collective nuclear rotation include examples with
a nonzero projection of the nuclear angular momentum onto the deformation
axis, and hence a Berry vector potential.  [There is also a scalar potential,
but it is independent of the slow variables,
and therefore at best could be observed in transitions
between instantaneous fast-variable eigenstates.]
Formally, the calculation of ${\cal I}$ appears to be higher order
in the slow velocity than the (linear) construction of the Berry vector
potential.  However, as the effect of that potential on motion of
the slow particles requires understanding of the kinetic energy for its
manifestation, the second-order terms surely are 
necessary
for a self-contained description of 
the geometry of adiabatic phenomena.   

Having acknowledged this principle, we still should note a quantitative
aspect which is so important that is tantamount to a qualitative 
distinction:  Unless the entire inertia tensor is generated adiabatically,
the adiabatic modification of the metric may be unobservably
small, being clearly of higher order 
in its effects than the Berry vector potential.
For example, in the LW case if the spinning particle has approximately
a Dirac gyromagnetic ratio, then the induced shift in the metric,
will be 

\vspace{-.15in}
\begin{equation}
\delta {\cal I}/{\cal I}
 = {\cal O}[2\pi{(\bf \nabla \hat B)^2/|B|}],
 \end{equation} where
here ${\bf |B|}$ is measured in units of an Aharonov-Bohm quantum of
flux. 
This quantity inevitably is much smaller than unity for any
reasonable setup.  That may well be the reason why its existence was overlooked
for so long before LW.  An open question is whether there exist systems
where the primitive and the induced contributions to the inertia are 
comparable, so that both must be taken into account for an accurate 
description of the motion.  
A promising place to look for such comparable contributions might be
the motion of quasiparticle excitations in a strongly correlated medium. 
Whatever the general answer to the question may be, the
induction in adiabatic processes 
of all conceivably relevant types of geometry (including ordinary
spatial geometry) appears inescapable.

\section{Full Newtonian adiabatic Hamiltonian}
Let us conclude with a comprehensive scheme for
computation of adiabatic quantum dynamics through second order in velocity.  
To weave together the discussion
here with previously identified elements, one begins with the computation of
the induced inertia, and uses the total $ \tilde{\cal I}={\cal I}_{\rm primitive}+{\cal I}_{\rm induced}$  to compute the inverse inertia $\tilde Q=\tilde {\cal I}^{-1}$.  An interesting point here is that in some cases there may be ambiguity about what is primitive and what is induced inertia, (for example, one might choose to redefine fast variables as describing, instead of motion with respect to a fixed frame, rather motion with respect to slow variables), but the sum should be unambiguous.  In terms of adiabatic perturbations, this statement 
seems quite natural:  The primitive inertia is simply an explicit (diagonal) second-order perturbation of the zero-velocity Hamiltonian, while the induced inertia comes from iterating a first-order off-diagonal perturbation.  By changing choices of basis one may shuffle contributions to the second-order diagonal part between primitive and induced.

 The total
adiabatic Hamiltonian in the context of the newer (induced-potential) stream of adiabatics 
was presented by Berry  \cite{ber}:
\begin{equation}
H_{eff}=V_{B-O}+(P-A_B)_iQ_{ij}(P-A_B)_j/2
+\hbar^{2}g_{ij}Q_{ij}/2, \label{comp}
\end{equation}
 where $V_{B-O}$ is the Born-Oppenheimer potential, 
including
all potential energies and also the kinetic energies
corresponding to fast degrees of freedom, averaged over those fast variables
for specified values of the slow variables.  The vector potential 
${\bf A}_B$ is
the connection associated with the Berry phase, which to this point in the present paper was kept 
hidden in the path-dependent transformation factor $U(t)$.  The scalar 
potential also comes from gradients of the path-dependent $U.$
 
  The effect of including induced inertia should be obvious at this point:  One rewrites 
  (\ref{comp}) as
  \begin{equation}
  H_{eff}=V_{B-O}+(P-A_B)_i\tilde Q_{ij}(P-A_B)_j/2
+\hbar^{2}g_{ij}\tilde Q_{ij}/2, 
\end{equation}
having exchanged the primitive and convention-dependent ${\cal I}$ for the complete quantity $\tilde {\cal I}$, hence replacing $Q$ by $\tilde Q = \tilde {\cal I}^{-1}$.

 Thus the Newtonian adiabatic Hamiltonian is determined by the ordinary geometry
(both primitive and induced) of the space of slow parameters,
 as well as by the induced or geometric vector and scalar
potentials, all in a coherent and consistent pattern.

\section{Outlook}

	While it  is worth recording the complete Newtonian adiabatic `package', the really interesting question is whether this package could provide any new insights into physical systems, and thus be something more than a mere catalogue entry.  The best prospect for such a development may be in analysis of strongly correlated systems, their  ground states and simple excitations.  Here is an analogy:  In classical electrodynamics, the hydrogen atom would be unstable against collapse, but quantum effects stabilize its ground state.  This makes the Newtonian approximation quite accurate for the ground-state structure.  
	
	Similarly, perhaps the exponentially suppressed jumps in adiabatic dynamics would simply disappear if one were using the adiabatic approximation to describe a stable structure, such as a many-body ground state, or a state built on that ground state with some fixed number of quasiparticles, each carrying a conserved charge.  	Again, the evident stability of these configurations suggests that the Newtonian description may become accurate, once one treats the adiabatic parameters as quantum variables.  In particular, for such an enterprise in the case of the fractional quantum Hall effect, where with interactions neglected there is no kinetic energy, induced inertia clearly becomes essential to the description.   

\section{Acknowledgments}	
	Michael Berry long ago pointed out important  references, including \cite{scalar} and \cite{lw}, and recently posed a crucial clarifying question.  Robert Littlejohn made illuminating remarks about the approach in \cite{lw}. Evan Fink made interesting comments and suggestions during a visit in summer 2002 as a Research Experiences for Undergraduates scholar.  This work was supported in part by the National Science Foundation, Grant  PHY-0140192.  Hospitality at the Newton Institute for Mathematical Sciences, Cambridge, UK in 1995 during  early thinking about this subject is much appreciated.

\end{document}